\documentclass[twocolumn,aps,pra]{revtex4}
\usepackage{epsfig}
\usepackage[english]{babel}
\usepackage{latexsym}
\usepackage{graphics}
\usepackage{subfigure}
\usepackage{epsfig}
\usepackage{graphicx}
\usepackage{dcolumn}
\usepackage{amsmath}
\usepackage{hyperref}
\usepackage{amssymb}
\usepackage{appendix}

\begin{document}

\title{Knee structure in the laser intensity dependence of high-order harmonic generation for graphene}
\author{Fulong Dong, Jie Liu$^{*}$}

\date{\today}

\begin{abstract}
We investigate the high-order harmonic generation (HHG) of graphene irradiated by linearly polarized lasers with intensities in a wide range from $10^{8}$ W/cm$^2$ to $10^{13}$ W/cm$^2$.
We find a striking knee structure in the laser intensity dependence of HHG, which consists of a linear growth regime, followed by a plateau of saturated HHG and then a transition to nonlinear growth.
The knee structure is rather universal for the various harmonic orders and has been certified by two-band density-matrix equation calculations as well as \textit{ab initio} time-dependent density functional theory calculations.
Based on the two-band model, we reveal the underlying mechanisms:
The linear growth behavior can be analytically depicted by the perturbative theory,
while the plateau of saturated HHG and the transition to nonlinear growth are caused by destructive and constructive quantum interference of harmonics generated by the electrons corresponding to the lattice momenta around Dirac points and M points in the Brillouin zone, respectively.
In particular, we find that tuning the Fermi energy can effectively alter the knee structure, while the profile of the knee structure is not sensitive to temperature.
Our calculations of the third-order harmonic versus tuned Fermi energy are compared with recent experiments, showing good agreement.
Our predicted knee structure and its associated properties are observable with the current experimental techniques.

\end{abstract}
\affiliation{Graduate School, China Academy of Engineering Physics, Beijing 100193, China}

\maketitle

\section{Introduction}
With the rapid development of laser technology, high-order harmonic generation (HHG) from the gases of atoms and molecules has been widely studied over the past several decades \cite{Corkum1,Lewenstein,Itatani,Meckel}, leading to the birth of a new branch of physics, namely, attosecond science \cite{Ferenc}.
More recently, much attention has been paid to HHG in solid materials \cite{Luu,Ghimire} due to the possibility of probing the electronic band structure \cite{Corkum3,Lanin} and obtaining coherent and bright attosecond pulses \cite{Luu,Garg}.

Graphene is a simple but specific two-dimensional material, in which there are only two carbon atoms per unit cell and the atoms are orderly arranged in a periodic hexagonal lattice.
It has important applications because of its unusual optical properties \cite{Castro}.
HHG in graphene has attracted much attention both experimentally \cite{Yoshikawa,Cox} and theoretically \cite{SCARZURR,YuntianZhang}.
HHG of graphene is found to be enhanced by elliptically polarized light \cite{Yoshikawa,Candong,Yongkang,Shunsuke}, in contrast to the atomic cases.
The harmonic ellipticity from graphene irradiated by elliptically \cite{SCARZURR,ZCHEN} and linearly \cite{ZCHEN,Fulong} polarized lasers has also been studied, and apparent elliptic 9th harmonic emission has been observed and thoroughly investigated under a linearly polarized laser.
In addition, HHG in bilayer and twisted bilayer graphene has also been studied \cite{MSMrudul,RuxinLi}.

In this work, we attempt to investigate the laser intensity dependence of HHG in graphene.
Note that in the situations of atoms and molecules, the issues associated with laser intensity dependence have been extensively addressed for single and double ionization \cite{Jieliu,BWalker,CGuo,CCornaggia,YHLai}, as well as HHG \cite{CGWahlstrom,LALompre}.
Atomic or molecular ionization is found to transition from multiphoton ionization to above-threshold ionization, tunnel ionization and finally over-the-barrier ionization with increasing laser intensity \cite{KAmini}.
Additionally, the HHG yield and "cutoff" of the harmonic photon energy are dramatically altered by the laser intensity.
In contrast to atomic and molecular systems, the HHG in graphene is generated by the sum current of the electrons with various lattice momenta in the Brillouin zone.
The quantum interference between the harmonics generated by different electrons will play an important role in the HHG.
Obviously, the laser intensity will affect the yields and especially the phases of the harmonics generated by these electrons, which are therefore expected to alter the structure in the profile of the laser intensity dependence of HHG in graphene.
Motivated by this consideration, in this paper, we calculate the harmonic yield of graphene for a wide range of laser intensities from $10^{8}$ W/cm$^2$ to $10^{13}$ W/cm$^2$ using both the two-band density-matrix equations and time-dependent density functional theory.
The simulation results of both methods demonstrate a striking knee structure that consists of a region of linear growth, then a plateau of saturated HHG, and a transition to nonlinear growth. The underlying mechanisms are analyzed.
In particular, we find that tuning the Fermi energy can effectively alter the knee structure, while the profile of the knee structure is not sensitive to temperature.

This paper is organized as follows.
We describe our calculation methods for the two-band density-matrix equations in the velocity gauge and time-dependent density functional theory in Sec. \ref{s2} and Sec. \ref{s3}, respectively.
Section \ref{s4} presents our main calculated results, and the mechanisms of the knee structure are discussed in Sec. \ref{s5}.
We also discuss the effects of finite temperature and Fermi energy tuning on the knee structure in Sec. \ref{s6}.
Finally, Sec. \ref{s7} presents our conclusion.

\section{Model}

\subsection{Two-band density-matrix equations}
\label{s2}

Graphene is a two-dimensional single layer of carbon atoms arranged in a honeycomb lattice \cite{Castro,PRWallace}, and its reciprocal space structure is a hexagonal lattice structure, as shown in Fig. 1(a).
The tight-binding Hamiltonian $H_{0}$ for electrons in
graphene considering that electrons can only hop to nearest-neighbor atoms has the form
\begin{eqnarray}
\begin{aligned}
H_{0}= \begin{pmatrix} 0 & \gamma_{0}f(\textbf{k}) \\ \gamma_{0} f^{*}(\textbf{k}) & 0 \end{pmatrix},
\end{aligned}
\end{eqnarray}
where \textbf{k} is the lattice momentum.
In our calculation, $\gamma_{0}=0.1$ a.u. is the nearest-neighbor hopping energy, and $f(\textbf{k})=e^{-i \textit{k}_{y}d}+2\cos(\sqrt{3}\textit{k}_{x}d/2)e^{-i\textit{k}_{y}d/2}$, with a carbon-carbon bond length of $d= 1.42$ \AA.
Diagonalization of the $H_{0}$ matrix can yield energy eigenvalues, which describe the valence ($v$) band $E_{v}(\textbf{k}) = -\gamma_{0} \vert f(\textbf{k}) \vert$ and conduction ($c$) band $E_{c}(\textbf{k}) = \gamma_{0} \vert f(\textbf{k}) \vert$.

Of particular importance for the physics of graphene are the two points $\textsc{K}$ and $\textsc{K}^{'}$ at the corners of the graphene Brillouin zone.
Their positions in momentum space are given by $\textsc{K}=(\dfrac{2\pi}{3\sqrt{3}d},\dfrac{2\pi}{3d})$ and $\textsc{K}^{'}=(\dfrac{4\pi}{3\sqrt{3}d},0)$, as shown in Fig. 1(a).
Figure 1(b) exhibits the electronic dispersion in the honeycomb lattice, where $E_{\textsc{F}}$ is the Fermi energy of doped graphene.

%%%%%%%%%%%%%%%%%%%%%%%%%%%%%%%%%%%%%%%%%%%%%%%%%%%%%
\begin{figure}[t]
\begin{center}
{\includegraphics[width=8.5cm, height=4cm]{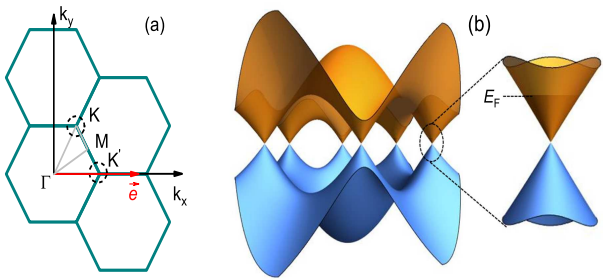}}
\caption{(a) Brillouin zone of the reciprocal lattice of graphene. Points $\textsc{K}$ and $\textsc{K}^{'}$ are two degenerate Dirac points, and the red arrow indicates the polarization direction of the electric field. (b) Dispersion relation of valence and conduction bands of graphene under the nearest-neighbor tight-binding approximation. $\textit{E}_{\textsc{F}}$ is the Fermi energy of doped graphene.}
\label{fig:graph1}
\end{center}
\end{figure}
%%%%%%%%%%%%%%%%%%%%%%%%%%%%%%%%%%%%%%%%%%%%%%%%%%%%%

We simulate the laser-graphene interaction by the two-band density-matrix equations (TBDMEs) in the velocity gauge.
Within the dipole approximation, they read
\begin{align}
i \dfrac{\partial}{\partial t}&\rho_{mn}(\textbf{k},t)=[E_{m}(\textbf{k})-E_{n}(\textbf{k})]\rho_{mn}(\textbf{k},t) \nonumber\\
&+\textit{\textbf{A}}(t) \cdot [\hat{\textbf{p}}(\textbf{k}),\hat{\rho}]_{mn} 
- i \Gamma_{i} \rho_{mn}(\textbf{k},t)(1-\delta_{mn})\nonumber\\
&-i \Gamma_{e} [\rho_{mn}(\textbf{k},t) - \rho_{mn}(\textbf{k},t=0)]\delta_{mn},
\end{align}
where $\hat{\rho}$ is the density matrix comprising elements $\rho_{mn}=\langle m,\textbf{k}\vert \rho \vert n,\textbf{k} \rangle$, where $m$ and $n$ represent the valence or conduction band.
$\Gamma_{i}$ and $\Gamma_{e}$, independent of $\textbf{k}$, are relaxation parameters.
%and they are set to zero unless mentioned otherwise.
The momentum matrix $\hat{\textbf{p}}(\textbf{k})$ consists of the intraband dipole elements $\textbf{p}_{cc}(\textbf{k})= \triangledown_{\textbf{k}} E_{c}(\textbf{k}) =- \textbf{p}_{vv}(\textbf{k})$ and the interband dipole elements
$\textbf{p}_{cv}(\textbf{k})=i (E_{c}(\textbf{k})-E_{v}(\textbf{k})) \textbf{D}_{cv}(\textbf{k}) =- \textbf{p}_{vc}(\textbf{k})$, where $\textbf{D}_{cv}(\textbf{k}) = i \langle u_{c,\textbf{k}}(\textbf{r})  \vert  \triangledown_{\textbf{k}} \vert u_{v,\textbf{k}}(\textbf{r}) \rangle$ and $u_{c,\textbf{k}}(\textbf{r})$ [$u_{v,\textbf{k}}(\textbf{r})$] is the periodic part of the Bloch wavefunction for the conduction (valence) band of graphene with crystal momentum $\textbf{k}$ \cite{GVampa,ShichengJ}.

$\textit{\textbf{A}}(t) = f(t) A_{0} \sin(\omega_{0} t) \vec{e}$ is the vector potential of the laser field, and $f(t)$ is an eight-cycle sine square envelope.
$A_{0}$ is the amplitude of the vector potential, and $\omega_{0}$ is the frequency of the laser field corresponding to wavelength $\lambda = 5500$ nm.
$\vec{e}$ is the unit vector along the laser polarization direction, as indicated by the red arrow in Fig. 1(a).

At the initial moment $t=0$, the density operator $\rho_{mn}(\textbf{k},t=0)$ is equal to $\delta_{mn} f_{m \textbf{k}}$, which characterizes the equilibrium occupation of single-particle states at finite temperature $T$ and Fermi energy $\textit{E}_{\textsc{F}}$ \cite{ChengSipe},
where
\begin{align}
f_{m \textbf{k}} = [1+e^{(E_{m}(\textbf{k})-\textit{E}_{\textsc{F}}) / (k_{B} T)} ]^{-1}
\end{align}
is the Fermi-Dirac distribution with Boltzmann's constant $k_{B}$.

Next, the total current can be evaluated by
\begin{align}
\textbf{\textit{j}}(t)=\sum_{\textbf{k}}\textbf{\textit{j}}_{\textbf{k}}(t)
\end{align}
and
\begin{align}
\textbf{\textit{j}}_{\textbf{k}}(t)=\rm{Tr} \lbrace \hat{\rho}[\hat{\textbf{p}}(\textbf{k})+\textit{\textbf{A}}(t)] \rbrace =\sum_{\emph{mn}}\textbf{p}_{\emph{mn}}(\textbf{k})\rho_{\emph{nm}}(\textbf{k},t)+\textit{\textbf{A}}(t),
\end{align}
where $\rm{Tr}$ denotes the trace.

The harmonic yield can be evaluated using
\begin{align}
H(\omega)=\omega^{2} \vert F(\omega)\vert^{2},
\end{align}
in which
\begin{align}
F(\omega)=\mathcal{T}_{F}[\vec{e} \cdot \textbf{\textit{j}}(t)],
\end{align}
where $\mathcal{T}_{F}$ denotes the Fourier transform.

\subsection{Time-dependent density functional theory}
\label{s3}

Since our density-matrix equations only consider two energy bands under the tight-binding approximation, we check our main results with time-dependent density functional theory (TDDFT) \cite{Ullrich},
Within the TDDFT framework, the evolution of the wavefunction is computed by propagating the Kohn-Sham equations:
\begin{eqnarray}
\begin{aligned}
i \frac{\partial}{\partial t} \psi_{n \textbf{k}}(\mathbf{r}, t)=\hat{H}_{\texttt{KS}}(\mathbf{r}, t)\psi_{n \textbf{k}}(\mathbf{r}, t),
\end{aligned}
\end{eqnarray}
where $\psi_{n \textbf{k}}(\mathbf{r}, t)$ is the Bloch state with band index $n$ and lattice momentum $\textbf{k}$. $\hat{H}_{\texttt{KS}}$ is the Kohn-Sham Hamiltonian given by
\begin{align}
\hat{H}_{\texttt{KS}}(\mathbf{r}, t) = \frac{1}{2}[\hat{\textbf{p}}+\boldsymbol{A}(t)]^{2}+V_{\texttt{KS}}(\mathbf{r}, t),
\end{align}
where $V_{\mathrm{KS}}(\mathbf{r},t)=V(\mathbf{r},t)+V_{\text {Hartree }}[\texttt{n}](\mathbf{r},t)+V_{\mathrm{xc}}[\texttt{n}](\mathbf{r},t)$ is the Kohn-Sham potential and $V_{\text {Hartree }}[\texttt{n}](\mathbf{r},t)$ is the Hartree potential. $\texttt{n}(\mathbf{r},t)=\sum_{n, \textit{\textbf{k}}}\left|\psi_{n \textit{\textbf{k}}}(\mathbf{r},t)\right|^{2}$ is the electron density. $V(\mathbf{r},t)$ represents the interaction between valence electrons and the ionic core and is modeled by norm-conserving pseudopotentials. $V_{\mathrm{xc}}[\texttt{n}](\mathbf{r},t)$ is the exchange-correlation potential accounting for all nontrivial many-body effects. Here, we apply the generalized gradient approximation in the Perdew-Burke-Ernzerhof parametrization.

In the calculation of the HHG in graphene, a $60 \times 60 \times 1 $ k-point mesh is used to sample the Brillouin zone, and the real-space spacing is $0.2$ \AA. The Octopus package \cite{Strubbe,Strubbe2} is employed to perform the simulations.

We compute the total electronic current $ \textit{\textbf{j}}(\mathbf{r},t)$ from time-evolved wavefunctions.
The harmonic yield can be evaluated using $H(\omega)=\omega^{2} \vert F(\omega)\vert^{2}$,
in which $F(\omega)=\mathcal{T}_{F}[\vec{e} \cdot \int d^{3} \mathbf{r} \textit{\textbf{j}}(\mathbf{r},t)]$.

%%%%%%%%%%%%%%%%%%%%%%%%%%%%%%%%%%%%%%%%%%%%%%%%%%%%%%%
\begin{figure}[t]
\begin{center}
{\includegraphics[width=8.5cm, height=12cm]{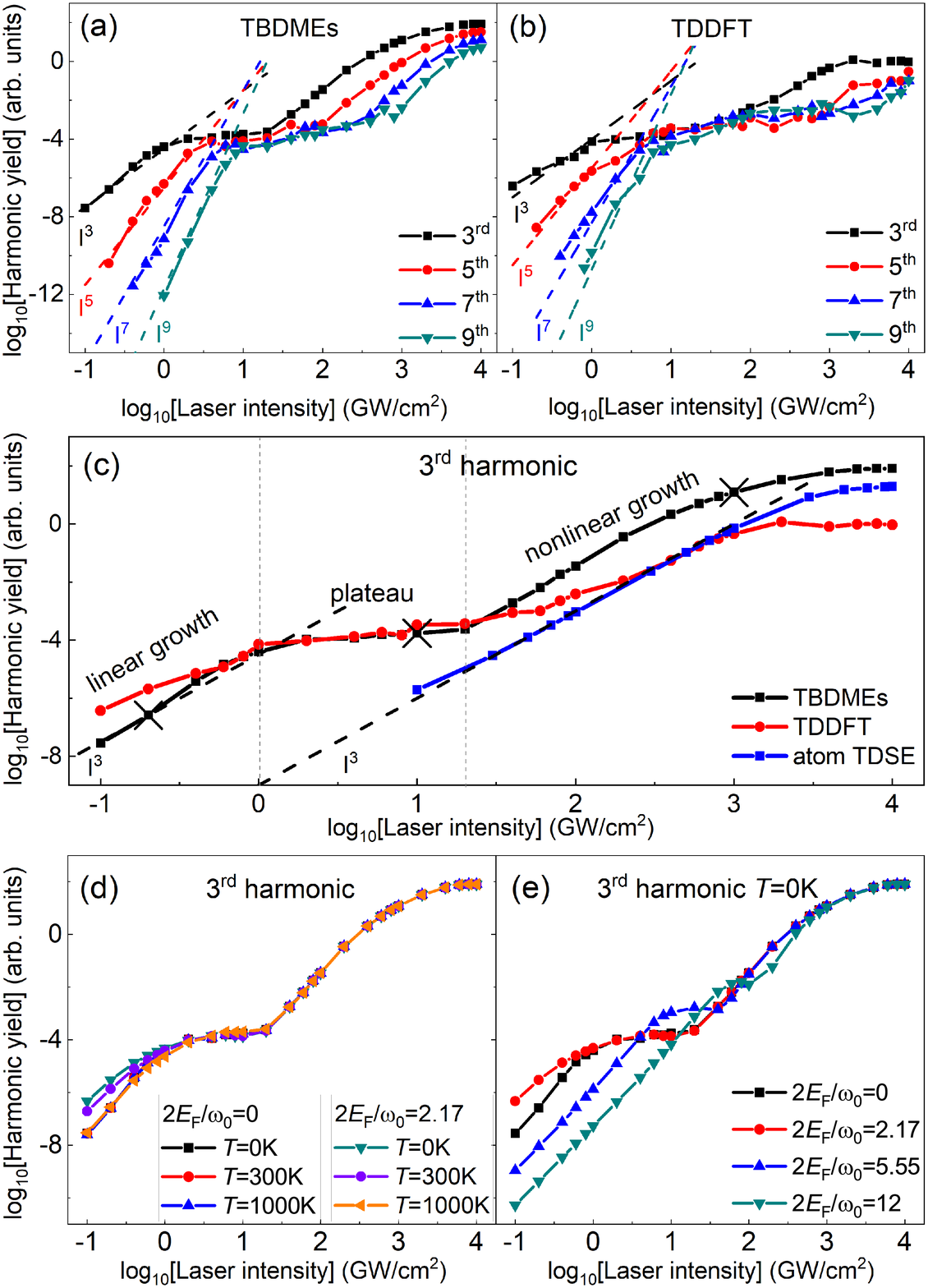}}
\caption{Laser intensity dependence of HHG calculated by the TBDMEs (a) and TDDFT (b).
(c) Comparison of the third harmonic yields of graphene calculated by the TBDMEs and TDDFT with that of the model atom calculated by the time-dependent Schr{\"o}dinger equation (TDSE).
The three crosses mark the laser intensities $I = 2 \times 10^{8}$W/cm$^{2}$, $1 \times 10^{10}$W/cm$^{2}$ and $1 \times 10^{12}$ W/cm$^{2}$.
In panels (a), (b) and (c), the dashed lines show the $I^{n}$ dependence for the $n$th harmonic.
(d) Effects of finite temperatures on the knee structure when the Fermi energy is tuned to $E_{\textsc{F}}  = 0$ or $2 E_{\textsc{F}}  = 2.17 \omega_{0}$.
(e) Effects of Fermi energy tuning on the knee structure when the temperature is zero.
Here, the relaxation parameters $\Gamma_{i}$ and $\Gamma_{e}$ are set to $0$ and the laser wavelength is $5500$ nm.}
\label{fig:graph1}
\end{center}
\end{figure}
%%%%%%%%%%%%%%%%%%%%%%%%%%%%%%%%%%%%%%%%%%%%%%%%%%%%%

\section{Knee structure in laser intensity-dependent HHG}

\subsection{Main results}
\label{s4}
In Fig. 2(a), we demonstrate the dependence of HHG calculated by the TBDMEs on the laser intensity over a wide range from $10^{8}$ W/cm$^2$ to $10^{13}$ W/cm$^2$ for the third- to ninth-order harmonics.
The results indicate that for each harmonic order, as the laser intensity increases, the harmonic yield first linearly increases, then saturates, and finally nonlinearly increases.
The profiles exhibit visible knee structures for harmonics up to the $9$th order.
To verify the TBDME results, we perform TDDFT calculations.
The results are shown in Fig. 2(b), which exhibit knee structures analogous to those in Fig. 2(a).
In Figs. 2(a) and 2(b), the dashed lines show that the $n$th-order harmonic yield $H(n \omega_{0})$ is proportional to the $n$th power of the laser intensity $I$, namely, $H(n \omega_{0}) \varpropto I^{n}$, as predicted by perturbation theory.

Note that the above interesting knee structure is not apparent in previous works that investigated the laser intensity dependence of HHG for crystalline solids and some two-dimensional materials \cite{Lanin,Yoshikawa,PeiyuXia,Hanzhe,LiangLi}. In these studies, their calculations were only implemented for a narrow range of laser intensities.

%%%%%%%%%%%%%%%%%%%%%%%%%%%%%%%%%%%%%%%%%%%%%%%%%%%%%%
\begin{figure*}[t]
\begin{center}
{\includegraphics[width=16cm,height=9cm]{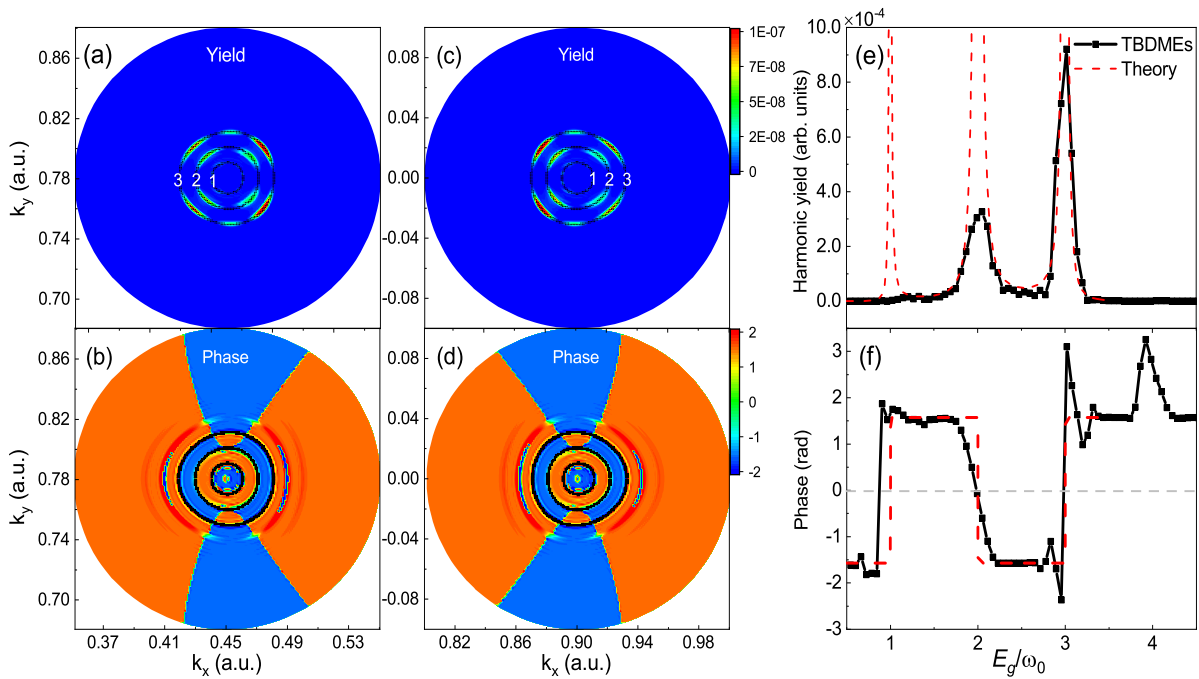}}
\caption{Third harmonic yield [(a) and (c)] and phase [(b) and (d)] calculated by Eq. (10) and Eq. (11), respectively, using the $\textbf{k}$ points around the $\textsc{K}$ [(a) and (b)] and $\textsc{K}^{'}$ [(c) and (d)] points of the Brillouin zone, as demonstrated by the two dashed circles in Fig. 1(a).
The black dot rings mark the positions where the energy gaps between the valence and conduction bands are $1 \omega_{0}$, $2 \omega_{0}$ and $3 \omega_{0}$.
(e) Third harmonic yield and (f) phase as a function of the energy gap.
In panels (e) and (f), the black square curves are numerically calculated by Eqs. (12) and (13) and the red dashed lines are evaluated by Eqs. (17) and (18) of perturbation theory, respectively.
Note that the above numerical results are calculated by the TBDMEs with Fermi energy $E_{\textsc{F}} =0$ and temperature $T=0 \textsc{K}$ for the laser intensity of $2\times 10^{8}$ W/cm$^{2}$ in the linear growth region.
}
\label{fig:graph1}
\end{center}
\end{figure*}
%%%%%%%%%%%%%%%%%%%%%%%%%%%%%%%%%%%%%%%%%%%%%%%%%%%%%%%%

%%%%%%%%%%%%%%%%%%%%%%%%%%%%%%%%%%%%%%%%%%%%%%%%%%%%%%%
\begin{figure*}[t]
\begin{center}
{\includegraphics[width=16cm,height=9cm]{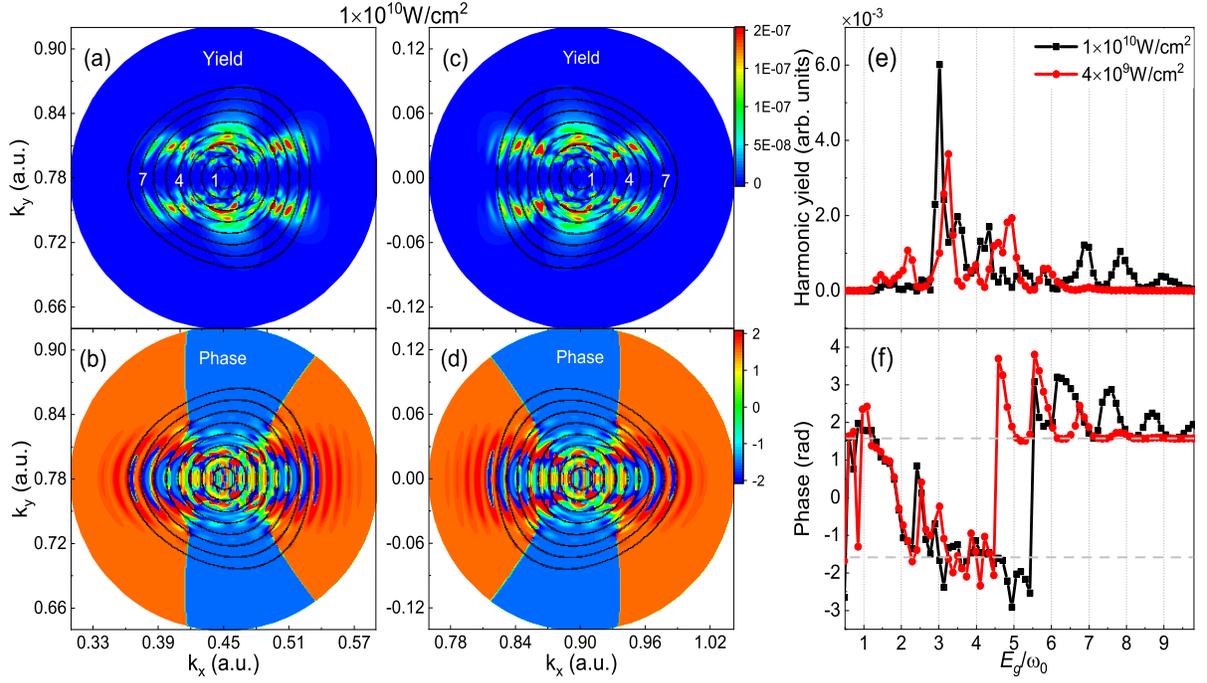}}
\caption{Panels (a), (b), (c) and (d) are the same as those in Fig. 3, but the laser intensity is $1 \times 10 ^{10}$ W/cm$^{2}$, which is in the plateau region.
Third harmonic yield (e) and phase (f) calculated by Eqs. (12) and (13) of the TBDMEs for laser intensities of $1 \times 10 ^{10}$ W/cm$^{2}$ (black square curves) and $4 \times 10 ^{9}$ W/cm$^{2}$ (red circle curves).}
\label{fig:graph1}
\end{center}
\end{figure*}
%%%%%%%%%%%%%%%%%%%%%%%%%%%%%%%%%%%%%%%%%%%%%%%%%%%%%%%%%%%

%%%%%%%%%%%%%%%%%%%%%%%%%%%%%%%%%%%%%%%%%%%%%%%%%%%%%%%%%%%%%
\begin{figure}[htb]
\begin{center}
{\includegraphics[width=8.5cm,height=6.75cm]{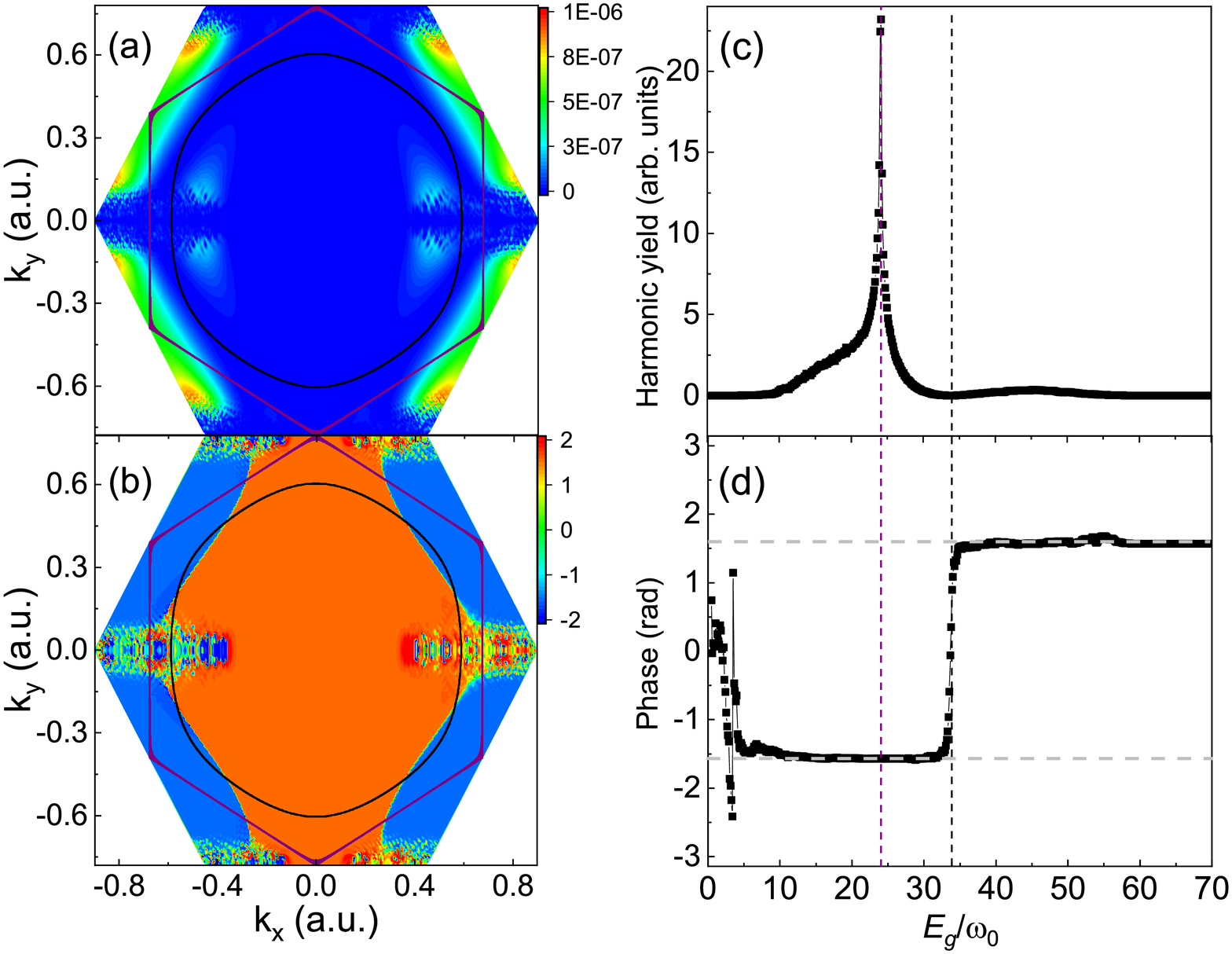}}
\caption{(a) Yield and (b) phase of the third harmonic generated by electrons corresponding to all $\textbf{k}$ points of the first Brillouin zone for the laser intensity of $1\times 10^{12}$ W/cm$^{2}$ in the nonlinear growth region.
(c) Third harmonic yield and (d) phase as a function of the energy gap.
In the left panels, the energy gaps of the $\textbf{k}$ points marked by purple and black curves are equal to the values indicated by the purple ($E_{g} = 24 \omega_{0}$) and black ($E_{g} = 34 \omega_{0}$) vertical dashed lines in the right panels, respectively.
}
\label{fig:graph1}
\end{center}
\end{figure}
%%%%%%%%%%%%%%%%%%%%%%%%%%%%%%%%%%%%%%%%%%%%%%%%%%%%%%%%%%%

Figure 2(c) compares the third harmonic yields of graphene calculated by the TBDME and TDDFT methods with that of the model atom.
The ionization potential of the model atom is $0.1$ a.u. \cite{MDFeit,Note}.
We find that the linear growth region, plateau region, and nonlinear growth region correspond to laser intensity ranges from $1 \times 10^{8}$ W/cm$^{2}$ to $1 \times 10^{9}$ W/cm$^{2}$, $1 \times 10^{9}$ W/cm$^{2}$ to $2 \times 10^{10}$ W/cm$^{2}$, and $2 \times 10^{10}$ W/cm$^{2}$ to $1 \times 10^{13}$ W/cm$^{2}$, respectively.
In contrast, the third harmonic yield of the model atom increases
monotonically, i.e., linearly, from $1 \times 10^{10}$ W/cm$^{2}$ to $1 \times 10^{12}$ W/cm$^{2}$. As the laser intensity increases further, the increase trend of the harmonic yield is suppressed when compared to the linear growth. 
No apparent knee structure is observed.
The phenomenon of a linear increase followed by nonlinear growth of the atomic HHG
has also been observed in many experiments on HHG \cite{CGWahlstrom,LALompre}.

Gate tuning of the Fermi energy allows control of the optical properties, leading to many promising applications in optoelectronics and photonics \cite{FengWang,FHKoppens}.
Additionally, the effects of finite temperature on HHG are often taken into account \cite{Alonso,TaoJiang,GianSoavi}. Here, we also attempt to consider the influence of the Fermi energy and finite temperature on the knee structure.
Figure 2(d) shows that the profile of the knee structure does not change with temperature when the Fermi energy is equal to $0$.
When the Fermi energy is tuned to $2 E_{\textsc{F}} = 2.17 \omega_{0}$, the linear growth region of the knee structure is weakly influenced by temperature.
In contrast, tuning the Fermi energy has a significant effect on this structure (see Fig. 2(e)) when the temperature is set to zero.
When the Fermi energy is tuned to $2 E_{\textsc{F}} = 2.17 \omega_{0}$, compared with $E_{\textsc{F}} = 0$, the harmonic yield in the linear growth region markedly increases.
When the Fermi energy is tuned to $2 E_{\textsc{F}} = 5.55 \omega_{0}$, the harmonic yield in the linear growth region is suppressed due to Pauli blocking, and that in the plateau region is enhanced.
Finally, the knee structure almost disappears when the Fermi energy is further increased to $2 E_{\textsc{F}} = 12 \omega_{0}$.
These results demonstrate that tuning the Fermi energy can effectively alter the knee structure, while the profile of the knee structure is not sensitive to temperature.

\subsection{Mechanism analysis at zero temperature}
\label{s5}

For simplicity, our following mechanism analysis is mainly focused on the third harmonic, but the conclusions are applicable to other harmonic orders.

\subsubsection{Linear region}

First, we investigate the mechanism of the linear growth region of the knee structure for the third harmonic.
We numerically calculate the yields [Figs. 3(a) and 3(c)] and phases [Figs. 3(b) and 3(d)] of the harmonic, which are generated by the electrons corresponding to lattice momenta $\textbf{k}$ around $\textsc{K}$ [Figs. 3(a) and 3(b)] and $\textsc{K}^{'}$ [Figs. 3(c) and 3(d)], as marked by the two dashed circles in Fig. 1(a).
The harmonic yield generated by the electron with lattice momentum $\textbf{k}$ is evaluated by
\begin{align} 
H^{\textbf{k}}(\omega)=\omega^{2} \vert F^{\textbf{k}}(\omega)\vert^{2},
\end{align}
with $F^{\textbf{k}}(\omega)=\mathcal{T}_{F}[\vec{e} \cdot \textbf{\textit{j}}^{\textbf{k}}(t)]$, and the phase $\phi^{\textbf{k}} (\omega)$ is calculated by
\begin{align} 
\tan[\phi^{\textbf{k}} (\omega)] = \dfrac{Im[ F^{\textbf{k}}(\omega)]}{Re[ F^{\textbf{k}}(\omega)]} .
\end{align}

In Figs. 3(a) to 3(d), the three black dot rings marked 1 to 3 correspond to energy gaps $E_{g} = 1 \omega_{0}$, $2 \omega_{0}$ and $3 \omega_{0}$, which are the energy differences between the valence and conduction bands.
As shown in Figs. 3(a) and 3(c), the electrons with both $E_{g} =2E_{c}(\textbf{k}) =  3 \omega_{0}$ and $E_{g} = 2E_{c}(\textbf{k}) =  2 \omega_{0}$ contribute to the third harmonic yield.

To clarify our results, we also numerically calculate the harmonic yield and phase as a function of energy gap $E_{g}$, as shown in Figs. 3(e) and 3(f), respectively.
The currents are calculated by $\textbf{\textit{j}}^{\textit{E}_{g}}(t)=\sum_{\textbf{k} \sim 2 \textit{E}_{c} (\textbf{k}) = \textit{E}_{g} } \textbf{\textit{j}}_{\textbf{k}}(t)$.
The harmonic yield is evaluated by
\begin{align} 
H^{\textit{E}_{g}}(\omega)=\omega^{2} \vert F^{\textit{E}_{g}}(\omega)\vert^{2},
\end{align}
with $F^{\textit{E}_{g}}(\omega)=\mathcal{T}_{F}[\vec{e} \cdot \textbf{\textit{j}}^{\textit{E}_{g}}(t)]$.
The phase $\phi^{\textit{E}_{g}} (\omega)$ is calculated by
\begin{align}
\tan[\phi^{\textit{E}_{g}} (\omega)]  = \dfrac{Im[ F^{\textit{E}_{g}}(\omega)]}{Re[ F^{\textit{E}_{g}}(\omega)]}.
\end{align}

In Fig. 3(e), our calculated results (black square curve) present two main harmonic yield peaks around $\textit{E}_{g} = 2 \omega_{0}$ and $3 \omega_{0}$, which are consistent with the results of Figs. 3(a) and 3(c).
In addition, on the two sides of each peak, the harmonic phase shifts by $\pi$, as shown in Fig. 3(f), which implies that the third harmonics generated by the electrons of different energy gaps interfere with each other.

These numerical results can be understood by the perturbation theory of graphene \cite{ChengSipe,Nonlinear,Vermeulen}.
Within the perturbation theory framework, the current can be calculated by $\textit{J}(t) = \sum_{n=1}^{\infty} \textit{J}^{(n)}(t)$, where $\textit{J}^{(n)}(t)$ is the $n$th-order perturbation expansion of the current.
The third-order current, which is related to Fermi energy $\textit{E}_{\textsc{F}} \geqslant 0$, is given by $\textit{J}^{(3)}(\frac{2 \textit{E}_{\textsc{F}}}{\omega},t) = \int d \omega \sigma^{(3)}(\frac{2 \textit{E}_{\textsc{F}}}{\omega},\omega) E^{3}(\omega) e^{-i 3 \omega t}$ \cite{Vermeulen}, where $\sigma^{(3)}$ is the third-order optical conductivity and $E(\omega)$ is the Fourier transform of electric field $E(t)$.
For simplicity, we consider $E(t) = E \cos(\omega_{0} t)$, and therefore, the third-order current can be evaluated by
\begin{align}
\textit{J}^{(3)}(\frac{2 \textit{E}_{\textsc{F}}}{\omega_{0}},t) =\sigma^{(3)}(\frac{2  \textit{E}_{\textsc{F}}}{\omega_{0}},\omega_0) E^{3} \cdot e^{-i 3 \omega_{0} t},
\end{align}
in which $\sigma^{(3)}(\dfrac{2 \textit{E}_{\textsc{F}}}{\omega_{0}},\omega_{0})  \varpropto i \dfrac{1}{\omega_{0}^4} T(\dfrac{\omega_{0}}{2 \textit{E}_{\textsc{F}}})$, as mentioned in Ref. \cite{Vermeulen}.
According to Eq. (6), the yield of the third harmonic is evaluated by
\begin{align}
&H^{(3)}(\frac{2 \textit{E}_{\textsc{F}}}{\omega_{0}})=(3 \omega_{0})^{2} \vert \mathcal{T}_{F}[\textit{J}^{(3)}(\frac{2 \textit{E}_{\textsc{F}}}{\omega_{0}},t)] \vert^{2} \nonumber\\
&= (3 \omega_0)^{2} \vert \sigma^{(3)}(\frac{2 \textit{E}_{\textsc{F}}}{\omega_{0}},\omega_{0}) \vert^{2} E^{6} \varpropto \frac{9 I^{3}}{\omega_{0}^{6}} \vert T(\dfrac{\omega_{0}}{2 \textit{E}_{\textsc{F}}}) \vert^{2},
\end{align}
where $T(x) = 17G(x) - 64G(2x) + 45G(3x)$, in which $G(x) = ln \vert \dfrac{1+x}{1-x} \vert + i \pi \theta (\vert x \vert -1)$ is a dimensionless complex function of real variable $x$. $\theta (y)$ is the Heaviside step function, equal to $0$ for $y < 0$ and $1$ for $y > 0$.

For a specific energy gap $E_{g}$, which can be related to the Fermi energy by $2\textit{E}_{\textsc{F}} = E_{g}$, the third-order current is
\begin{align}
\textit{J}^{E_{g}(3)}(t) = \lim _{\Delta \varepsilon \rightarrow 0} [ \textit{J}^{(3)}(\dfrac{E_{g} - \Delta \varepsilon}{\omega_{0}},t)
- \textit{J}^{(3)}(\dfrac{E_{g} + \Delta \varepsilon}{\omega_{0}},t) ].
\end{align}
The corresponding harmonic yield is evaluated by
\begin{align} 
&H^{\textit{E}_{g}(3)}=(3 \omega_0)^{2} \vert F^{\textit{E}_{g}(3)} \vert^{2} \nonumber\\
&\varpropto (3 \omega_0)^{2} I^{3} \left| \lim _{\Delta \varepsilon \rightarrow 0} [ \sigma^{(3)}(\dfrac{E_{g} - \Delta \varepsilon}{\omega_{0}},t) -  \sigma^{(3)}(\dfrac{E_{g} + \Delta \varepsilon}{\omega_{0}},t) ] \right|^{2}, 
\end{align}
with $F^{\textit{E}_{g}(3)}=\mathcal{T}_{F}[J^{\textit{E}_{g}(3)}(t)]$.
The phase $\phi^{\textit{E}_{g}(3)}$ is calculated by
\begin{align}
\tan[\phi^{\textit{E}_{g}(3)}]  = \dfrac{Im[ F^{\textit{E}_{g}(3)}]}{Re[ F^{\textit{E}_{g}(3)}]}.
\end{align}

In Figs. 3(e) and 3(f), the yield and phase demonstrated by the red dashed lines are calculated by Eq. (17) and Eq. (18), respectively.
The calculated results qualitatively agree with the analytical results.
In addition, according to Eq. (15), the third harmonic yield can be evaluated by $H^{(3)}(\frac{2 \textit{E}_{\textsc{F}}}{\omega_{0}}) =  (3 \omega_0)^{2} \vert \sigma^{(3)}(\frac{2 \textit{E}_{\textsc{F}}}{\omega_{0}},\omega_{0}) \vert^{2} E^{6}  \propto  (3 \omega_0)^{2} \vert \sigma^{(3)}(\frac{2 \textit{E}_{\textsc{F}}}{\omega_{0}},\omega_{0}) \vert^{2} I^{3}$, which implies that the third harmonic yield is proportional to the third power of the laser intensity.

Equation (17) indicates that the third harmonic yield generated by electrons corresponding to a specific energy gap is also proportional to the third power of the laser intensity.
The third harmonics generated by electrons of different energy gaps may interfere with each other due to phase differences.

\subsubsection{Plateau region}
Compared with the results for the linear growth region, Figs. 4(a) and 4(c) show that for $1 \times 10^{10}$ W/cm$^{2}$ in the plateau region, more electrons whose lattice momenta are also around Dirac points of the Brillouin zone (corresponding to the energy gap range from $E_{g} = 0$ to $E_{g} = 9 \omega_{0}$) play roles in the generation of the third harmonics.
The phases are more disordered, as shown in Figs. 4(b) and 4(d).

%%%%%%%%%%%%%%%%%%%%
\begin{figure*}[hbpt]
\begin{center}
{\includegraphics[width=17cm,height=6cm]{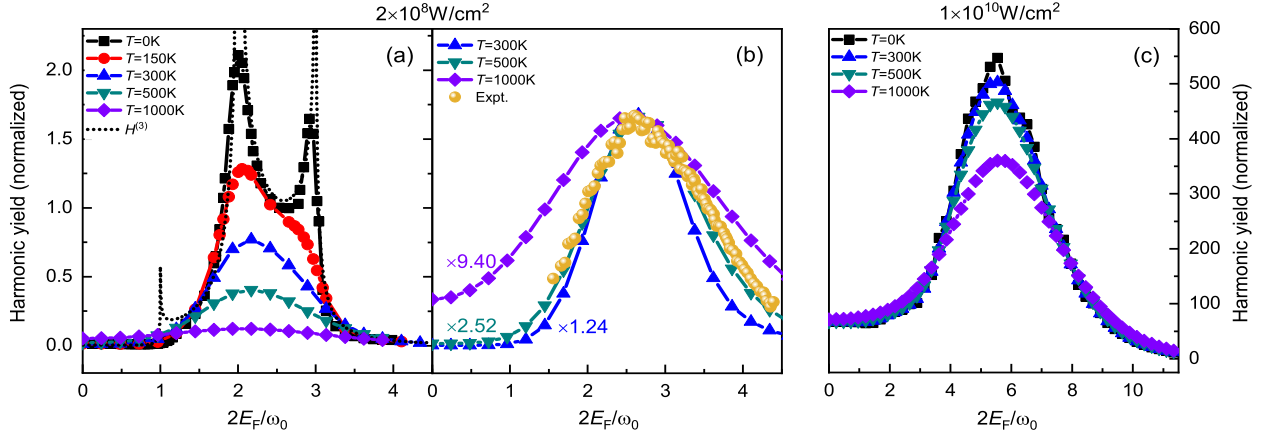}}
\caption{Third harmonic yield as a function of Fermi energy $\textit{E}_{\textsc{F}}$ at different temperatures for $2 \times 10^{8}$ W/cm$^{2}$ (a) in the linear growth region and $1 \times 10^{10}$ W/cm$^{2}$ (c) in the plateau region. The dashed line in (a) represents the perturbation theory result of the third harmonic calculated by Eq. (15). (b) Comparison between experimental results of Ref. \cite{Alonso} and our numerical results for $2 \times 10^{8}$ W/cm$^{2}$, in which the relaxation parameters of $\Gamma_{i} =0 $ and $\Gamma_{e} = 0.002$ a.u. have been taken into account in the TBDMEs.
}
\label{fig:graph1}
\end{center}
\end{figure*}
%%%%%%%%%%%%%%%%%%%%%%%%%%%%%%

Figures 4(e) and 4(f) show the dependence of the yield and phase of the third harmonic on the energy gap $E_{g}$.
For the laser intensity of $1 \times 10^{10}$ W/cm$^{2}$ (black square curves), when the energy gap is below $3 \omega_{0}$, the yields of the third harmonic are weak, and the phases are disordered; therefore, the contribution of electrons in this energy gap range to the harmonic yield can be ignored.
In the energy gap range from $3 \omega_{0}$ to $4.5 \omega_{0}$, strong third harmonics are generated, and the phases of the harmonics are near $-\pi/2$.
Finally, two yield peaks appear at energy gaps of $7 \omega_{0}$ and $8 \omega_{0}$, and the phases around these energy gaps are near $\pi/2$.
For the two energy gap ranges (from $3 \omega_{0}$ to $4.5 \omega_{0}$ and from $7 \omega_{0}$ to $8 \omega_{0}$), the $\pi$ phase difference may cause destructive interference of the harmonics.

To demonstrate the generation mechanism of the plateau region more clearly, we also demonstrate the results for $4 \times 10^{9}$W/cm$^{2}$ for comparison.
In the plateau region, as the laser intensity increases, the harmonic yield related to the energy gap does not rapidly increase.
For a higher laser intensity of $1 \times 10^{10}$ W/cm$^{2}$, the electrons located in a wider energy gap range contribute to the generation of the harmonics, but the destructive quantum interference mentioned above may restrain the harmonic yield.
This is the generation mechanism of the plateau region.

\subsubsection{Nonlinear growth}
For the laser intensity of $1 \times 10^{12}$ W/cm$^{2}$, Figs. 5(a) and 5(b) present the yield and phase of the third harmonic generated by electrons corresponding to all lattice momenta $\textbf{k}$ of the first Brillouin zone.
According to Fig. 5(a), in contrast to the linear growth region and the plateau region, in the nonlinear growth region, the harmonic yield is dominated by the electrons whose lattice momenta are near $\textsc{M}$ points.
Figures 5(c) and 5(d) exhibit the harmonic yield and phase as a function of energy gap $E_{g}$.
In the left panels of Fig. 5, the energy gaps of the $\textbf{k}$ points marked by purple and black curves are equal to the values indicated by purple ($E_{g} = 24 \omega_{0}$) and black ($E_{g} = 34 \omega_{0}$) vertical dashed lines, respectively.

In Fig. 5(d), a phase transition from $-\pi /2$ to $ \pi /2$ appears at the energy gap $E_{g} = 34 \omega_{0}$.
Although the third harmonics are also generated by the electrons located in the energy gap range of $E_{g} \geqslant 34 \omega_{0}$, the harmonic yield is very weak, and therefore, we mainly focus on the harmonics generated by the electrons of $E_{g} < 34 \omega_{0}$.
In this range, Fig. 5(c) exhibits an obvious yield peak located at $E_{g} = 2 E_{c}(\textbf{k}) = 24 \omega_{0} \approx 0.2$ a.u., as indicated by the purple vertical dashed line.
The origin of this peak can be understood from two aspects.
First, the density of states located at $E_{c}(\textbf{k}) \approx 0.1$ a.u., which corresponds to $\textsc{M}$ points, is the largest in the Brillouin zone \cite{Castro}.
Second, as marked by the purple curve in Fig. 5(b), the phases of the third harmonics are all $\pi /2$.
Hence, the yield peak appears due to constructive quantum interference of the harmonics.

Note that in the range of laser intensities from $3 \times 10^{12}$ W/cm$^2$ to $1 \times 10^{13}$ W/cm$^2$ the harmonic yield of the model atom presents a  growth trend similar to the nonlinear growth region of graphene. However, underlying  mechanism is quite different: When  the laser intensity further increases, during the rise of the electric field,  the electron has tunnelled out from the barrier formed by the Coulomb potential and the laser field, and they can not recombine with its parent nucleus to emit high harmonics. This will lead to somehow saturation in HHG for atomic system \cite{Corkum1}.

\subsection{Effects of finite temperature and Fermi energy tuning}
\label{s6}

For the laser intensity of $2 \times 10^{8}$ W/cm$^{2}$, Fig. 6(a) shows that when the temperature is 0, the third harmonic yield significantly changes with increasing Fermi energy, and a double-peak structure located at $2 E_{\textsc{F}} = 2 \omega_{0}$ and $2 E_{\textsc{F}} = 3 \omega_{0}$ is observed.
The result agrees with the dashed line, which is the prediction from perturbation theory for graphene calculated by Eq. (15).
As the temperature increases, the double-peak structure is smoothed and evolves into a single-peak structure centered at $2 E_{\textsc{F}} = 2.17 \omega_{0}$.

When the temperature is set to 0, compared with $E_{\textsc{F}} = 0$, the third harmonic yield of doped graphene with $2 E_{\textsc{F}} = 2.17 \omega_{0}$ is significantly enhanced.
The influence of this Fermi energy tuning on the knee structure is clearly shown by the comparison between the black and red curves of Fig. 1(e).
When the Fermi energy is $0$, Fig. 6(a) shows that the third harmonic yields are not sensitive to temperature.
When the Fermi energy is tuned to $2 E_{\textsc{F}} = 2.17 \omega_{0}$, the harmonic yield decreases with increasing temperature.

In Fig. 6(b), we compare our numerical results in which the relaxation parameters of $\Gamma_{i} = 0$ and $\Gamma_{e} = 0.002$ a.u. (corresponding to a relaxation time of $12.1$ fs) have been taken into account with the experimental results of Ref. \cite{Alonso}.
For three typical temperatures $T = 300 \textsc{K}$, $500 \textsc{K}$ and $1000 \textsc{K}$, the numerical results have been scaled by factors of $1.24$, $2.52$ and $9.4$, respectively.
The numerical results for the temperature of $500 \textsc{K}$ match the experimental results, showing good agreement.

For the $1 \times 10^{10}$ W/cm$^{2}$ intensity in the plateau region, the dependence of the third harmonic yield on the Fermi energy exhibits a single-peak structure centered at $2 E_{\textsc{F}} = 5.5 \omega_{0}$, as shown in Fig. 6(c).
Compared with the results for $2 \times 10^{8}$W/cm$^{2}$, the single-peak structure is more insensitive to temperature.
More importantly, for temperature $T = 0$, the third harmonic yield of doped graphene with $2 E_{\textsc{F}} = 5.5 \omega_{0}$ is 10 times stronger than that with $E_{\textsc{F}} = 0$.
This result is also reflected in the comparison between the blue triangle curve and black square curve in Fig. 1(e), where the harmonic yield in the plateau region increases with the Fermi energy transition.
At the same time, the harmonic yield in the linear growth region is obviously suppressed due to Pauli blocking.
These results indicate that for the linear growth region and plateau region, tuning of the Fermi energy can significantly influence the knee structure.

\section{Conclusion}
\label{s7}
In summary, we have investigated the HHG of graphene irradiated by linearly polarized lasers with various intensities and found a striking knee structure, in contrast to the atomic and molecular situations.
The underlying mechanism is determined to be the destructive and constructive quantum interference of harmonics generated by the electrons corresponding to the lattice momenta around Dirac points and M points in the Brillouin zone, respectively.
Our findings have also been certified by \textit{ab initio} TDDFT calculations.
In particular, we find that tuning the Fermi energy can effectively alter the knee structure, while the profile of the knee structure is not sensitive to temperature.
The knee structure is rather universal, and its associated properties can be observed with the current experimental techniques.
In fact, our calculations of the third-order harmonic versus tuned Fermi energy have been compared with a recent experiment, showing good agreement.

\section*{ACKNOWLEDGMENTS}

This work is supported by NSAF (Grant No. U1930403). We acknowledge valuable discussions with Professor Difa Ye.


\begin{thebibliography}{2}
\bibitem{Note} In our work, the one-dimensional time-dependent Schr{\"o}dinger equation of the model atom in the length gauge is numerically solved by the spectral method \cite{MDFeit}.
Here, we use the soft-Coulomb potential $V(x)=-Z/\sqrt{\varepsilon+x^{2}}$ and set the effective charge $Z=0.172$ atomic units as well as the smoothing parameter $\varepsilon=0.5$ to ensure that the ionization potential of the model atom is $0.1$ atomic units, which is comparable to the energy of the valence band of graphene.
Additionally, the frequency of the electric field we use is $0.057$ atomic units.
\end{thebibliography}
\end{document}